\documentstyle[epsbox,12pt]{article}

\begin{document}

{\bf\Large Feasibility of GRB with TeV gamma ray all sky monitor}

Satoko Osone
{\small
{\it  Matsudo 1083-6 A-202, Matsudo, Chiba 271-0092, Japan}\\
}

\section*{Abstract}

We discuss feasibility of Gamma ray burst (GRB) with TeV gamma ray all sky monitor and discuss necessity of TeV gamma ray cherenkov all sky monitor.
 
\section{Introduction}
Tibet II(Amenomori et al., 1996), HEGRA AIROBICC(Padilla et al., 1998) suggest 10-20 TeV emission  from GRB.
Milagrito detected TeV emission from GRB 970417a (Atkins et al., 2003).
This luminosity is almost same with  that of low energy range.
There are models of TeV emission through proton synchrotron emission (Totani 1999) or Synchrotron Self Compton (SSC) of electron (eg. Dermer, Chiang \& Mitman 2000; Pilla \& Loeb 1998), from GRB.
Totani (1999) discuss small number of detection of GRB is attributed to a number of low redshift GRB because gamma ray emission from high redshift GRBs attenuate by making an interaction with infrared background. 
Therefore, Tibet III and Milagro(Atkins et al., 2001) aim to observe GRB with low TeV gamma ray energy where the interaction is less.
Results of Tibet III and Milagro have not been presented yet. 
Here, we propose more sensitive detector, TeV cherenkov all sky monitor (IACT ASM).
We discuss feasibility of GRB with Tibet III, Milagro and IACT ASM and discuss necessity of IACT ASM.
 
\section{TeV gamma ray all sky monitor}
There are two cosmic ray detectors, Milagro and TibetIII at low TeV energy.
We propose IACT ASM which is more sensitive detector.
We use Frenel lens instead of reflecting mirrors, which cover Field of View (F.O.V.) of 60$\times$60 degrees with an accuracy 0.1 degree.
We consider two cases, IACT ASM of diameter 4m, 10m.
As for sensitivity of IACT ASM (4m), we refer to observation result of Cangaroo I which diameter is 3.8 m and angular resolution 0.1 degree.
Yoshikoshi (1996) observed  Vela pulsar with Cangaroo I. Flux is obtained  2.5$\times10^{-12}$ s$^{-1}$ cm$^{-2}$  at 2.5 TeV with  174h. on observation(4.4$\sigma$).
With 1 sec observation, we obtain sensitivity of 2.0$\times 10^{-9}$ s$^{-1}$cm$^{-2}$(4.4$\sigma$). 
We use a formula S/N$ = SAT/\sqrt{BAT\Omega}$ ($S$: sensitivity(ph s$^{-1}$ cm$^{-2}$), $B$: Background(ph s$^{-1}$ cm$^{-2}$ str$^{-1}$) , $A$: effective area(cm$^2$), $T$:integrated time(sec), $\Omega$:angular resolution(str)).
This value is not correct because of image parameter cut, but good indicator.

Next, we derive a sensitivity of Tibet III which is sensitive to gamma ray above 3 TeV.
With flux of cosmic ray  $10^{-6}$s$^{-1}$ cm$^{-2}$ and an angular resolution 1 degree, we obtain a sensitivity of 10$^{-8}$s$^{-1}$cm$^{-2}$ (4$\sigma$) with 1 sec observation.

\begin{table}
\caption{The comparison of Tibet III and IACT ASM 4m which are sensitive to gamma ray above 3 TeV. Time duration of GRB 1 sec is considered.}
\begin{tabular}{ccc}\hline
  & Tibet III & IACT ASM 4m  \\ \hline
effective area(cm$^2$) & 10$^8$ & 10$^8$ \\
Sensitivity(s$^{-1}$cm$^{-2}$) & 10$^{-8}$ &  10$^{-9}$ \\
photon limit(s$^{-1}$cm$^{-2}$) & 10$^{-7}$ & 10$^{-7}$ \\
F.O.V. &   45$\times$45 deg.($\pi^2/16$ rad)&  60$\times$60 deg.($\pi^2/9$ rad)\\
duty & 100 \%  &  10 \%\\
detection rate of GRB & 0.1/5 yr  & 0.02/5yr   \\ \hline
\end{tabular}
\end{table}

\section{Feasibility of GRB}

Here, we assume a luminosity at TeV gamma ray energy range is as same as that at low energy range detected with BATSE,  which is suggested from previous result(Atkins et al.,2003).

\begin{figure}[htb]
 \begin{center}
\psbox[height=8.5cm,width=8.5cm]{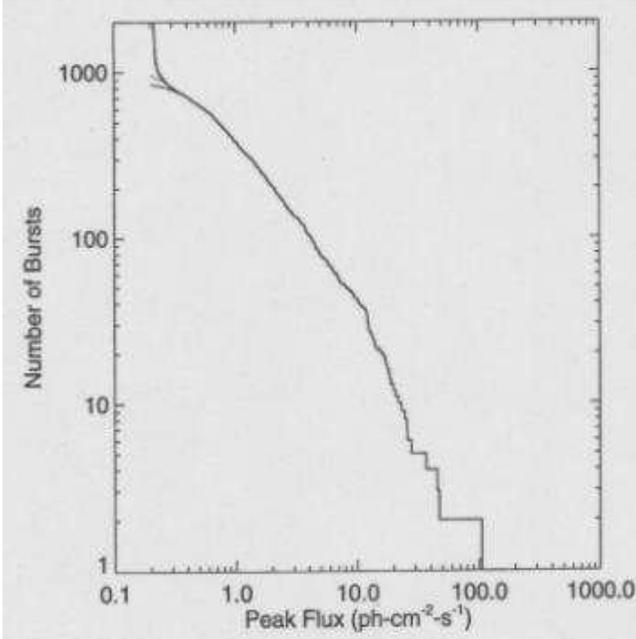}

 \end{center}
\caption{The relation between integrated number $N$ above some sensitivity $S$ and sensitivity $S$ with BATSE(Pacieasa et al. 1999). The energy range is 50-300keV and threshold of time duration is 1024 msec. Data is taken between 1991 Apr. and 1996 Aug..}
\end{figure}

\begin{figure}[htb]
 \begin{center}
\psbox[height=8.5cm,width=8.5cm]{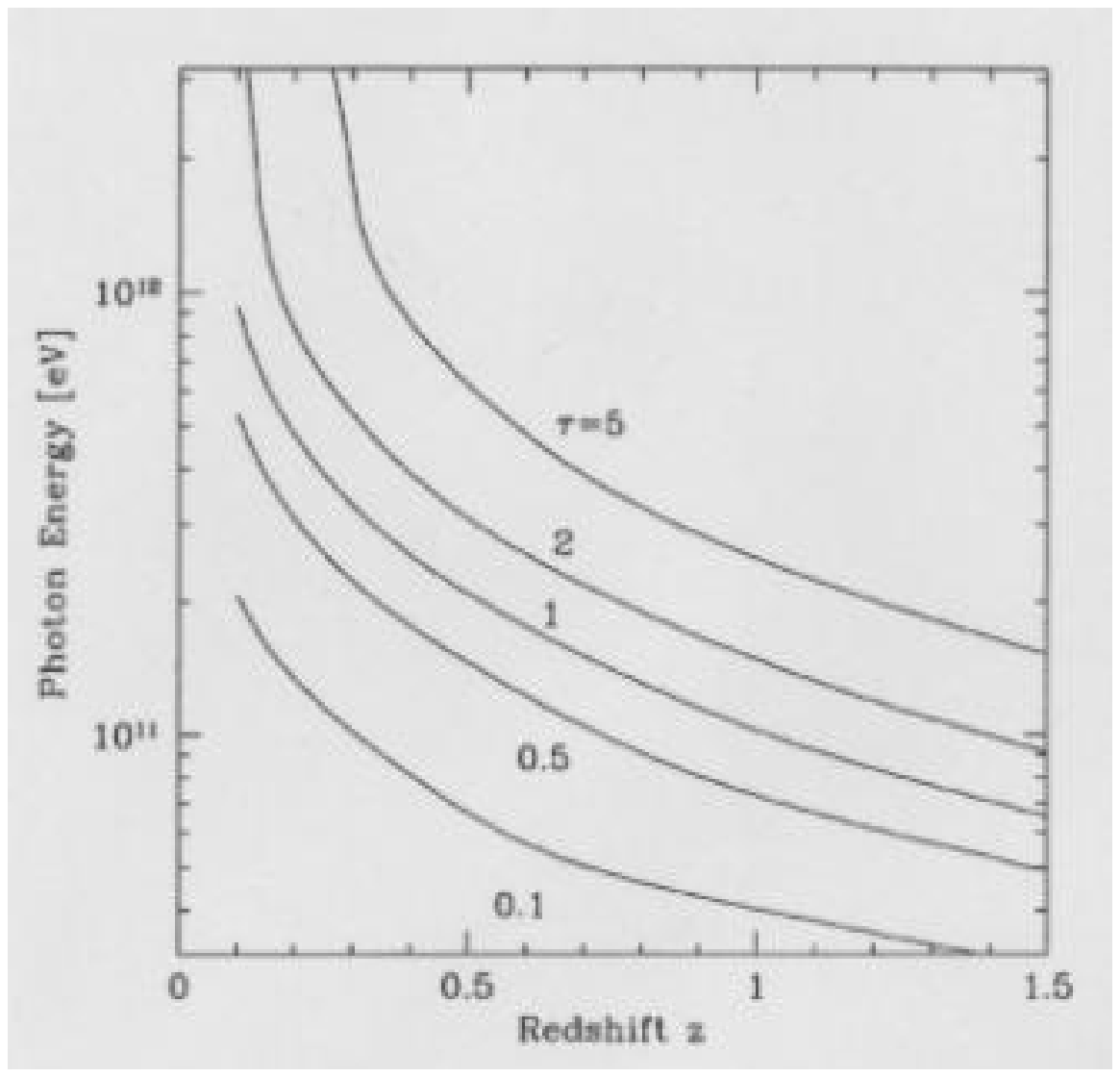}
 \end{center}
\caption{The contour map of an intergalactic optical depth of very high energy gamma rays , as a function of observed photon energy and source redshift(Totani 2000).}
\end{figure}

We show a relation between integrated number $N$ above some sensitivity $S$ and sensitivity $S$(logN$-$logS)(Pacieasa et al. 1999) with BATSE at figure 1.
The peak energy in energy spectra $vF_v$ (erg/s) vs. $v$(Hz) is 300 keV(Preece et al. 2000).
Therefore, the maximum flux 100 s$^{-1}$cm$^{-2}$ at 300 keV as shown figure 1 corresponds to 1$\times10^{-5}$ s$^{-1}$cm$^{-2}$ at 3 TeV.
Here, there is a photon limit which we need more than 10 photons(3 $\sigma$) from a source.
We found that we can not observe flux below $10^{-7}$s$^{-1}$cm$^{-2}$ for typical time duration 1 sec and an effective area $10^8$cm$^2$.
We found there is no merit for IACT ASM 4m because there is a photon limit.

There is an interaction between infrared background and TeV gamma rays.
At 3 TeV, we can observe attenuated gamma rays up to optical depth $\tau$ 5, where flux of $10^{-5}$s$^{-1}$cm$^{-2}$ become $10^{-7}$s$^{-1}$cm$^{-2}$ (photon limit).
Therefore, we can detect GRB within redshift $z$ 0.3 in figure 2.
Number of GRB within $z$ 0.3 is 1/1000 of all (figure 1 in Totani 2000).
A number of observation of GRB with BATSE is 1637/5yr.
BATSE exposure fraction is 0.5(Pacieasa et al. 1999) and number of detection rate of GRB is 3274/5yr for all sky.
We show possible detection rate of GRB with both IACT ASM 4m and Tibet III at table 1.

For detectors which is sensitive to 100 GeV, we consider only photon limit.
Maximum flux 100 s$^{-1}$cm$^{-2}$ at 300 keV as shown figure 1 corresponds to 3$\times10^{-4}$ s$^{-1}$cm$^{-2}$ at 100 GeV.
Time duration of GRB ranges from 0.1 to 100sec(Paciesas et al. 1999). We consider two time durations of GRB, 1, 100 sec.
We assume F.O.V of 45 degree for Milagro.
We can observe flux 1/30 of photon limit with 100 sec time duration for Milagro.
We derive rate of  GRB above flux 3 s$^{-1}$cm$^{-2}$ at 300 keV as 100/5yr from figure 1 and calculate number of GRB as 200/5yr for all sky, 6/5yr for F.O.V. of Milagro. 
We show possible detection rate of GRB with both Milagro and IACT ASM 10 m in table 2.

\begin{table}
\caption{Comparison bwteeen Milagro and IACT ASM 10m which are sensitive to gamma ray above 100 GeV. Two time duration 1 sec and 100 sec are considered. }
\begin{tabular}{ccc}\hline
& Milagro & IACT ASM 10 m \\ \hline
effective area(cm$^2$) & 10$^4$ & 10$^8$ \\
photon limit(1sec)(s$^{-1}$cm$^{-2}$) & 10$^{-3}$  & 10$^{-7}$ \\
photon limit(100sec)(s$^{-1}$cm$^{-2}$) & 10$^{-5}$  & 10$^{-9}$ \\
duty & 100 \% & 10 \% \\
detection rate of GRB(1 sec) & 0  & $\ge$18/5yr \\
detection rate of GRB(100 sec) & 6/5yr    & $\ge$18/5yr \\ \hline
\end{tabular}
\end{table}

We found an important parameter is not sensitivity, but effective area for detection of GRB.
We found a feasibility of GRB is high with IACT ASM 10m.
We propose IACT ASM of 10 m with Frenel lens.

We thank assi. prof Takita for information of Tibet III.
We thank Dr. Totani and Dr. Tamagawa for information of GRB.
We thank Dr. Kawasaki for information of Frenel lens.
We thank assi. prof Mori for information of IACT 10m.

\section{References}

\begin{itemize}
\setlength{\itemsep}{-1.5mm}
\setlength{\itemindent}{-8mm}
\item[]Amenomori et al., 1996, A\&A, 311, 919.
\item[]Atkins et al., astro-ph/0110513
\item[]Atkins et al., 2003, ApJ, 583, 824.
\item[]Dermer, Chiang \& Mitman, 2000, ApJ,537, 785.
\item[]Paciesas et al., 1999, ApJS, 122,465.
\item[]Padilla et al., 1998, A\&A, 337, 43.
\item[]Pilla \& Loeb, 1998, ApJ, 494, L167.
\item[]Preece et al., 2000, ApJS, 126, 19.
\item[]Totani.T, 1999, Astropart., 11,451.
\item[]Totani T., 2000, ApJ, 536, L23.
\item[]Yoshikoshi T., 1996, Doctor thesis(Tokyo Institute of Technology) 
\end{itemize}

\end{document}